# Double multiple-relaxation-time lattice Boltzmann model for solid-liquid phase change with natural convection in porous media

Qing Liu, Ya-Ling He

*Key Laboratory of Thermo-Fluid Science and Engineering of Ministry of Education, School of Energy and Power Engineering, Xi'an Jiaotong University, Xi'an, Shaanxi, 710049, China*

**Abstract**

In this paper, a double multiple-relaxation-time lattice Boltzmann model is developed for simulating transient solid-liquid phase change problems in porous media at the representative elementary volume scale. The model uses two different multiple-relaxation-time lattice Boltzmann equations, one for the flow field and the other for the temperature field with nonlinear latent heat source term. The model is based on the generalized non-Darcy formulation, and the solid-liquid phase change interface is traced through the liquid fraction which is determined by the enthalpy method. The model is validated by numerical simulations of conduction melting in a semi-infinite space, solidification in a semi-infinite corner, and convection melting in a square cavity filled with porous media. The numerical results demonstrate the efficiency and accuracy of the present model for simulating transient solid-liquid phase change problems in porous media.

## 1. Introduction

Solid-liquid phase change phenomena in saturated porous media have gained considerable research attention due to the importance in a wide variety of natural and engineering systems. Applications include the thermal energy storage, welding and casting of a manufacturing process, melting and freezing of soils, artificial freezing of ground for construction and mining purposes, freezing of soil around the heat exchanger coils of a ground based heat pump, and so on [1-3]. A better understanding of solid-liquid phase change process in porous media is quite necessary. However, solutions of solid-liquid phase change problems are difficult to obtain because of the inherent nonlinearities caused by the formation and propagation of the moving phase change interface, as well as the associated computational and physical complexities. Analytical results to such problems are limited to the cases with extremely simple geometry and boundary conditions. Therefore numerical methods are more frequently employed in modeling solid-liquid phase change problems in porous media. Over the past several decades, various traditional numerical methods, such as the finite volume method, the finite difference method, and the finite element method, have been employed to study solid-liquid phase change problems with convection heat transfer in porous media based on the discretization of some semiempirical models (e.g., the Darcy model, the Brinkman-extended Darcy model, and the Forchheimer-extended Darcy model) [1-7].

The lattice Boltzmann (LB) method, as a mesoscopic numerical method originates from the lattice-gas automata (LGA) method [8], has been developed into an attractive numerical technique to simulate fluid flows and model physics in fluids [9-19]. Unlike the traditional numerical methods, which are based on the discretization of macroscopic equations, the LB method is based on mesoscopic kinetic equations for particle distribution functions. Owing to its kinetic background and distinctive computational features, the LB method is a very attractive numerical technique to solve nonlinear macroscopic systems [20]. In recent years, the LB method has been successfully applied to simulate solid-liquid phase change problems. Generally, the existing LB models for solid-liquid phase change problems can be classified into two categories, i.e., the phase-field method [21-24] and the enthalpy-based method [25-35]. In the phase-field method, the solid-liquid phase change interface is traced by the order parameter and is of finite thickness. To reproduce the dynamics of the phase change interface equation, the phase-field method requires extremely finer lattice spacing to resolve the phase change interface region in principle. On the other hand, in the enthalpy-based method, the position of the phase change interface is tracked through the liquid fraction which is calculated by the total enthalpy, and finer lattice spacing in the phase change interface region is not necessary. Because of its conceptual simplicity and computational efficiency, the enthalpy-based method has been widely employed in LB simulations of solid-liquid phase change problems with or without considering fluid flow. Among the above mentioned enthalpy-based LB models, Huber et al. [28] proposed an enthalpy-based LB model for solving conduction and convection melting problems. Huber and co-workers' model can be used to study convection melting in porous media at the pore scale. In Ref. [31], Gao and Chen proposed an LB model for simulating melting with natural convection in porous media at the representative elementary volume (REV) scale based on the local thermal equilibrium model. Recently, Gao et al. [35] developed a three-distribution-function thermal LB model for natural convection with solid-liquid phase change in porous media at the REV scale under the local thermal non-equilibrium condition.

It is emphasized that the existing LB models for solid-liquid phase change problems in porous media at the REV scale are based on the Bhatnagar-Gross-Krook (BGK) collision operator [36], which is the most commonly employed collision operator in the LB method. However, the BGK collision operator directly results in several inherent deficiencies of the BGK-LB model, such as the numerical instability at low viscosity and the inaccuracy in treating boundary conditions [37-42]. These deficiencies of the BGK model can be addressed by using the multiple-relaxation-time (MRT) model proposed by d'Humières [43]. In the LB community, it has been widely accepted that the MRT model is superior over its BGK counterpart in terms of numerical stability and accuracy [41, 42]. In addition, the MRT model has sufficient tunable parameters to cover the anisotropic diffusion coefficient tensor [44]. These features make the MRT-LB model more useful in investigating practical problems. Hence, the aim of this paper is to develop an MRT-LB model for simulating transient solid-liquid phase change problems with natural convection through porous media at the REV scale. In the model, the Brinkman-Forchheimer-extended Darcy model (also called the generalized non-Darcy model) is employed to describe the momentum transfer in porous media, and the solid-liquid phase change interface is traced through the liquid fraction which is determined by the enthalpy method. The rest of this paper is organized as follows. The macroscopic governing equations are described in Section 2. In Section 3, the MRT-LB model for solid-liquid phase change problems with natural convection in porous media is presented in detail. The numerical tests and some discussions are given in Section 4. Finally, Section 5 concludes the paper.

## 2. Macroscopic governing equations

It is assumed that the porous matrix and the phase change material (PCM) are isotropic and homogeneous, the porous matrix and the solid phase of the PCM are rigid, and the local thermal equilibrium model is applicable. In addition, the physical properties of the solid and liquid phases are assumed to be constant except the density in the buoyancy term of the momentum equation, which is treated according to the Boussinesq approximation. Furthermore, it is assumed that the density change during the phase change process can be neglected, and the flow is incompressible without viscous heat dissipation. Based on the above mentioned assumptions, the dimensional governing equations for solid-liquid phase change with convection heat transfer in porous media at the REV scale can be described as follows [1]:

$$\nabla \cdot \mathbf{u} = 0, \tag{1}$$

$$\frac{\partial \mathbf{u}}{\partial t} + (\mathbf{u} \cdot \nabla)\left(\frac{\mathbf{u}}{\varphi}\right) = -\frac{1}{\rho_0}\nabla(\varphi p) + \upsilon_e \nabla^2 \mathbf{u} + \mathbf{F}, \tag{2}$$

$$\sigma \frac{\partial T}{\partial t} + \mathbf{u} \cdot \nabla T = \nabla \cdot (\alpha_e \nabla T) - \phi \frac{L_a}{c_{pl}} \frac{\partial f_l}{\partial t}, \tag{3}$$

where $\mathbf{u}$, $p$, and $T$ are the volume-averaged velocity, pressure and temperature, respectively, $\rho_0$ is the mean density of the PCM, $\upsilon_e$ is the effective viscosity, $\alpha_e$ is the effective thermal diffusivity, $\sigma$ is the thermal capacity ratio, $L_a$ is the latent heat of solid-liquid phase change, $c_{pl}$ is the liquid specific heat, $\phi$ is the porosity of the porous media, $f_l$ is the liquid fraction of the PCM in the pore space, and $\varphi$ is the liquid fraction of the PCM in the volume element ($\varphi = \phi f_l$). $\mathbf{F} = (F_x, F_y)$ denotes the total body force due to the presence of a porous medium and other external force fields, which is given by [45, 46]

$$\mathbf{F} = -\frac{\varphi \upsilon_l}{K}\mathbf{u} - \frac{\varphi C_F}{\sqrt{K}}|\mathbf{u}|\mathbf{u} + \varphi \mathbf{G}, \tag{4}$$

where $\upsilon_l$ is the viscosity of the fluid (not necessarily the same as $\upsilon_e$), $K$ is the permeability, and $C_F$ is the inertial coefficient, $|\mathbf{u}| = \sqrt{u_x^2 + u_y^2}$, in which $u_x$ and $u_y$ are the components of the velocity $\mathbf{u}$ in the $x$- and $y$-directions, respectively. Based on the Boussinesq approximation, the buoyancy force $\mathbf{G}$ is given by

$$\mathbf{G} = g\beta(T - T_0)\mathbf{j}, \tag{5}$$

where $g$ is the gravitational acceleration, $\beta$ is the thermal expansion coefficient, $T_0$ is the reference temperature, and $\mathbf{j}$ is the unit vector in the $y$-direction. The inertial coefficient $C_F$ and the permeability $K$ of liquid flow through a porous medium with randomly distributed spherical particles can be obtained by [1, 2]

$$C_F = \frac{1.75}{\sqrt{175\varphi^3}}, \quad K = \frac{\varphi^3 d_m^2}{175(1-\varphi)^2}, \tag{6}$$

where $d_m$ is the mean diameter of the solid particle. The thermal capacity ratio $\sigma$ is the ratio between the mean thermal capacity of the mixture and the liquid thermal capacity

$$\sigma = \frac{\phi\rho\left[f_l c_{pl} + (1-f_l)c_{ps}\right] + (1-\phi)(\rho c_p)_m}{\rho c_{pl}}, \tag{7}$$

where the subscripts $l$, $s$ and $m$ refer to the properties of the liquid and solid phases of the PCM, and the porous matrix, respectively, and $c_p$ is the specific heat at constant pressure. The effective thermal diffusivity $\alpha_e$ is defined as $\alpha_e = k_e/(\rho_l c_{pl})$, in which $k_e$ is the effective thermal conductivity. Generally speaking, the effective thermal conductivity $k_e$ depends on the structure of the porous matrix as well as on the volume fractions and thermal conductivities of each constituent. For a porous medium with randomly distributed spherical inclusions, the effective thermal conductivity $k_e$ may be approximated by the following equation [47]

$$k_e + \phi\left(\frac{k_m - k_p}{k_p^{1/3}}\right)k_e^{1/3} - k_m = 0. \tag{8}$$

where $k_p = f_l k_l + (1-f_l)k_s$ is the mean thermal conductivity of the PCM.

Using the following dimensionless variables:

$$X = \frac{x}{H}, \quad Y = \frac{y}{H}, \quad \mathbf{U} = \frac{\mathbf{u}H}{\alpha_l}, \quad P = \frac{pH^2}{(\rho v \alpha)_l}, \quad \theta = \frac{T - T_0}{\Delta T}, \quad Fo = \frac{t\alpha_l}{H^2},$$

$$\lambda = \frac{\alpha_e}{\alpha_l}, \quad St = \frac{c_{pl}\Delta T}{L_a}, \quad Pr = \frac{\upsilon_l}{\alpha_l}, \quad Ra = \frac{g\beta\Delta T H^3}{\upsilon_l \alpha_l}, \quad Da = \frac{K}{H^2}, \quad J = \frac{\upsilon_e}{\upsilon_l}, \tag{9}$$

the following dimensionless governing equations can be obtained:

$$\nabla \cdot \mathbf{U} = 0, \tag{10}$$

$$\frac{1}{\varphi Pr}\left[\frac{\partial \mathbf{U}}{\partial Fo}+(\mathbf{U}\cdot\nabla)\left(\frac{\mathbf{U}}{\varphi}\right)\right]=-\nabla P+\frac{J}{\varphi}\nabla^2\mathbf{U}-\left(\frac{1}{Da}+\frac{C_F}{Pr\sqrt{Da}}|\mathbf{U}|\right)\mathbf{U}+Ra\theta\mathbf{j}, \qquad (11)$$

$$\sigma\frac{\partial \theta}{\partial Fo}+\mathbf{U}\cdot\nabla\theta=\nabla\cdot(\lambda\nabla\theta)-\frac{\phi}{St}\frac{\partial f_l}{\partial Fo}, \qquad (12)$$

where $Fo$ is the Fourier number (dimensionless time), $\lambda$ is the ratio of the effective thermal diffusivity to the liquid thermal diffusivity, $St$ is the Stefan number, $Pr$ is the Prandtl number, $Ra$ is the Rayleigh number, $Da$ is the Darcy number, $J$ is the viscosity ratio, $H$ is the characteristic length, and $\Delta T$ is the temperature difference (characteristic temperature).

## 3. MRT-LB model for solid-liquid phase change in porous media

In this section, a thermal MRT-LB model for solid-liquid phase change in porous media is developed based on the enthalpy method. In the model, an incompressible D2Q9 MRT-LB equation is employed to solve the flow field, and a D2Q5 MRT-LB equation is proposed to solve the temperature field with nonlinear latent heat source term. In order to determine the position of the solid-liquid phase change interface, the enthalpy formulation is incorporated into the model by adding a source term into the MRT-LB equation of the temperature field.

### 3.1 MRT-LB equation for the flow field

For the flow field, the MRT-LB equation with an explicit treatment of the forcing term can be written in general as the following [40, 48, 50]:

$$\mathbf{f}(\mathbf{x}+\mathbf{e}\delta_t, t+\delta_t)-\mathbf{f}(\mathbf{x}, t)=-\mathbf{M}^{-1}\mathbf{\Lambda}\left[\mathbf{m}(\mathbf{x}, t)-\mathbf{m}^{(eq)}(\mathbf{x}, t)\right]+\mathbf{M}^{-1}\delta_t\left(\mathbf{I}-\frac{\mathbf{\Lambda}}{2}\right)\mathbf{S}, \qquad (13)$$

where $\mathbf{M}$ is a $9\times 9$ orthogonal transformation matrix, $\mathbf{\Lambda}=\mathbf{M}\tilde{\mathbf{\Lambda}}\mathbf{M}^{-1}$ is a non-negative $9\times 9$ diagonal relaxation matrix ($\tilde{\mathbf{\Lambda}}$ is the collision matrix), and $\mathbf{I}$ is the identity matrix. The boldface symbols, $\mathbf{f}$, $\mathbf{m}$, $\mathbf{m}^{(eq)}$, and $\mathbf{S}$, are 9-dimensional column vectors:

$$\mathbf{f}(\mathbf{x}+\mathbf{e}\delta_t, t+\delta_t)=\left(f_0(\mathbf{x}+\mathbf{e}_0\delta_t, t+\delta_t),\ldots,f_8(\mathbf{x}+\mathbf{e}_8\delta_t, t+\delta_t)\right)^{\mathrm{T}},$$

$$\mathbf{f}(\mathbf{x},t) = \left(f_0(\mathbf{x},t), f_1(\mathbf{x},t), \ldots, f_8(\mathbf{x},t)\right)^\mathrm{T},$$

$$\mathbf{m}(\mathbf{x},t) = \left(m_0(\mathbf{x},t), m_1(\mathbf{x},t), \ldots, m_8(\mathbf{x},t)\right)^\mathrm{T},$$

$$\mathbf{m}^{(eq)}(\mathbf{x},t) = \left(m_0^{(eq)}(\mathbf{x},t), m_1^{(eq)}(\mathbf{x},t), \ldots, m_8^{(eq)}(\mathbf{x},t)\right)^\mathrm{T},$$

$$\mathbf{S} = \left(S_0, S_1, \ldots, S_8\right)^\mathrm{T}, \qquad (14)$$

where T denotes the transpose operator, $f_i(\mathbf{x},t)$ is the volume-averaged density distribution function with a discrete velocity $\mathbf{e}_i = (e_{ix}, e_{iy})$ at position $\mathbf{x} = (x,y)$ and time $t$, $\mathbf{m}(\mathbf{x},t)$ and $\mathbf{m}^{(eq)}(\mathbf{x},t)$ are the velocity moments and the corresponding equilibrium moments, respectively, and $S_i$ is the component of the forcing term $\mathbf{S}$. The nine discrete velocities of the D2Q9 model (see Fig. 1) are given by [49]

$$\mathbf{e}_i = \begin{cases} (0,0), & i=0 \\ \left(\cos\left[(i-1)\pi/2\right], \sin\left[(i-1)\pi/2\right]\right)c, & i=1\sim 4 \\ \left(\cos\left[(2i-9)\pi/4\right], \sin\left[(2i-9)\pi/4\right]\right)\sqrt{2}c, & i=5\sim 8 \end{cases}, \qquad (15)$$

where $c = \delta_x/\delta_t$ is the lattice speed, in which $\delta_t$ and $\delta_x$ are the discrete time step and lattice spacing, respectively. In the MRT model, the lattice speed $c$ is usually set to be 1, which leads to $\delta_x = \delta_t$.

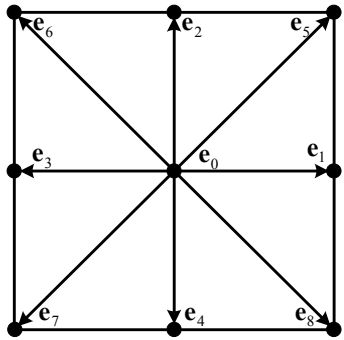

Fig. 1. Discrete velocities of the D2Q9 model.

The transformation matrix $\mathbf{M}$ linearly transforms the discrete distribution functions $\mathbf{f}$ in the velocity moment $\mathbf{V} = \mathbf{R}^9$ to their velocity moments $\mathbf{m}$ in the moment space $\mathbf{M} = \mathbf{R}^9$:

$$\mathbf{m} = \mathbf{M}\mathbf{f} = (m_0, m_1, \ldots, m_8)^\mathrm{T}, \quad \mathbf{f} = \mathbf{M}^{-1}\mathbf{m} = (f_0, f_1, \ldots, f_8)^\mathrm{T}. \tag{16}$$

Considering the effect of the forcing term, the nine velocity moments for the D2Q9 model are given by

$$\mathbf{m} = \left(\rho, e, \varepsilon, j_x - \frac{\delta_t}{2}\rho F_x, q_x, j_y - \frac{\delta_t}{2}\rho F_y, q_y, p_{xx}, p_{xy}\right)^\mathrm{T}, \tag{17}$$

where $m_0 = \rho$ is the mass density, $m_1 = e$ is the second-order moment related to energy, $m_2 = \varepsilon$ is the fourth-order moment related to energy square, $m_{3,5} = j_{x,y}$ are components of the flow momentum $\mathbf{J} = (j_x, j_y) = \rho\mathbf{u}$, $m_{4,6} = q_{x,y}$ are the third-order moments related to energy flux, and $m_{7,8} = p_{xx,xy}$ are the second-order moments related to the diagonal and off-diagonal components of the stress tensor, respectively. With the ordering of the above velocity moments, the transformation matrix $\mathbf{M}$ can be constructed via the Gram-Schmidt orthogonalization procedure ($c = 1$) [37]

$$\mathbf{M} = \begin{pmatrix} 1 & 1 & 1 & 1 & 1 & 1 & 1 & 1 & 1 \\ -4 & -1 & -1 & -1 & -1 & 2 & 2 & 2 & 2 \\ 4 & -2 & -2 & -2 & -2 & 1 & 1 & 1 & 1 \\ 0 & 1 & 0 & -1 & 0 & 1 & -1 & -1 & 1 \\ 0 & -2 & 0 & 2 & 0 & 1 & -1 & -1 & 1 \\ 0 & 0 & 1 & 0 & -1 & 1 & 1 & -1 & -1 \\ 0 & 0 & -2 & 0 & 2 & 1 & 1 & -1 & -1 \\ 0 & 1 & -1 & 1 & -1 & 0 & 0 & 0 & 0 \\ 0 & 0 & 0 & 0 & 0 & 1 & -1 & 1 & -1 \end{pmatrix}. \tag{18}$$

Note that the row vectors of $\mathbf{M}$ are mutually orthogonal, i.e., $\mathbf{M}\mathbf{M}^\mathrm{T}$ is a diagonal matrix, thus the inverse of $\mathbf{M}$ can be obtained according to the formula: $\mathbf{M}^{-1} = \mathbf{M}^\mathrm{T}(\mathbf{M}\mathbf{M}^\mathrm{T})^{-1}$. In the system of $\{f_i\}$, the conserved quantities are only the density $\rho$ and the two components of the momentum $\mathbf{J} = \rho\mathbf{u}$. The other velocity moments are non-conserved quantities and as suggested by Lallemand and Luo [37], they relax linearly towards the corresponding equilibrium moments. The equilibrium moments $\{m_i^{(\mathrm{eq})} \mid i = 0, 1, \ldots, 8\}$ corresponding to the velocity moments $\{m_i \mid i = 0, 1, \ldots, 8\}$ are defined as [50]

$$\mathbf{m}^{(eq)} = \begin{pmatrix} \rho \\ -2\rho + 3\rho_0 |\mathbf{u}|^2/\varphi \\ \rho - 3\rho_0 |\mathbf{u}|^2/\varphi \\ \rho_0 u_x \\ -\rho_0 u_x \\ \rho_0 u_y \\ -\rho_0 u_y \\ \rho_0 \left(u_x^2 - u_y^2\right)/\varphi \\ \rho_0 u_x u_y/\varphi \end{pmatrix}. \tag{19}$$

For incompressible fluids considered in this work, the so-called incompressible approximation [51] has been employed in the above equilibrium moments, i.e., the density $\rho = \rho_0 + \delta\rho \approx \rho_0$ ($\delta\rho$ is the density fluctuation), and $\mathbf{J} \approx \rho_0 \mathbf{u}$. To include the influence of the porous medium, the porosity has been incorporated into the equilibrium moments.

The evolution process of the MRT-LB equation (13) consists of two steps, i.e., the collision step and streaming step [37]. The collision step is executed in the moment space:

$$\mathbf{m}^+(\mathbf{x}, t) = \mathbf{m}(\mathbf{x}, t) - \boldsymbol{\Lambda}\left[\mathbf{m}(\mathbf{x}, t) - \mathbf{m}^{(eq)}(\mathbf{x}, t)\right] + \delta_t\left(\mathbf{I} - \frac{\boldsymbol{\Lambda}}{2}\right)\mathbf{S}, \tag{20}$$

while the streaming step is still carried out in the velocity space:

$$f_i\left(\mathbf{x} + \mathbf{e}_i \delta_t, t + \delta_t\right) = f_i^+(\mathbf{x}, t), \tag{21}$$

where $\mathbf{f}^+(\mathbf{x}, t) = \mathbf{M}^{-1}\mathbf{m}^+(\mathbf{x}, t)$.

To get the correct equations of hydrodynamics, the components of the forcing term $\mathbf{S}$ in the moment space are defined as follows [50]

$$S_0 = 0, \quad S_1 = \frac{6\rho_0 \mathbf{u} \cdot \mathbf{F}}{\varphi}, \quad S_2 = -\frac{6\rho_0 \mathbf{u} \cdot \mathbf{F}}{\varphi}, \quad S_3 = \rho_0 F_x, \quad S_4 = -\rho_0 F_x,$$

$$S_5 = \rho_0 F_y, \quad S_6 = -\rho_0 F_y, \quad S_7 = \frac{2\rho_0\left(u_x F_x - u_y F_y\right)}{\varphi}, \quad S_8 = \frac{\rho_0\left(u_x F_y + u_y F_x\right)}{\varphi}. \tag{22}$$

The diagonal relaxation matrix $\boldsymbol{\Lambda}$ is given by

$$\boldsymbol{\Lambda} = \text{diag}(s_1, s_2, s_3, s_4, s_5, s_6, s_7, s_8)$$

$$= \text{diag}(1, s_e, s_\varepsilon, 1, s_q, 1, s_q, s_\upsilon, s_\upsilon). \tag{23}$$

where $\{s_i | i = 0, 1, \ldots, 8\}$ are dimensionless relaxation rates and $s_i \in (0, 2)$ for the non-conserved velocity moments.

The macroscopic fluid density $\rho$ and velocity $\mathbf{u}$ are defined as

$$\rho = \sum_{i=0}^{8} f_i, \tag{24}$$

$$\rho_0 \mathbf{u} = \sum_{i=0}^{8} \mathbf{e}_i f_i + \frac{\delta_t}{2} \rho_0 \mathbf{F}. \tag{25}$$

It should be noted that the total body force $\mathbf{F}$ also contains the velocity $\mathbf{u}$, so Eq. (25) is a nonlinear equation for the velocity $\mathbf{u}$. According to Ref. [52], the macroscopic velocity $\mathbf{u}$ can be calculated explicitly via

$$\mathbf{u} = \frac{\mathbf{v}}{d_0 + \sqrt{l_0^2 + l_1 |\mathbf{v}|}}, \tag{26}$$

where $\mathbf{v}$ is defined as

$$\mathbf{v} = \sum_{i=0}^{8} \mathbf{e}_i f_i / \rho_0 + \frac{\delta_t}{2} \varphi \mathbf{G}. \tag{27}$$

The two parameters $l_0$ and $l_1$ in Eq. (26) are given by

$$l_0 = \frac{1}{2}\left(1 + \varphi \frac{\delta_t}{2} \frac{\upsilon_l}{K}\right), \quad l_1 = \varphi \frac{\delta_t}{2} \frac{C_F}{\sqrt{K}}. \tag{28}$$

The macroscopic fluid pressure $p$ is defined as $p = c_s^2 \rho / \varphi$, and the effective kinetic viscosity $\upsilon_e$ is defined by $\upsilon_e = c_s^2 (\tau_\upsilon - 0.5) \delta_t$ with $s_7 = s_8 = s_\upsilon = 1/\tau_\upsilon$, where $\tau_\upsilon$ is the dimensionless relaxation time of the BGK model, and $c_s = c/\sqrt{3}$ is the sound speed of the D2Q9 model. Through the Chapman-Enskog analysis [48, 53] of the MRT-LB equation (13) in the moment space, the generalized Navier-Stokes equations (1) and (2) can be recovered in the incompressible limit under the assumption that the flow varies slowly in time, i.e., $t_c \gg L_c/c_s$ ($t_c$ and $L_c$ are the characteristic time and length of the system, respectively). It should be noted that, when all the relaxation rates $\{s_i | i = 0, 1, \ldots, 8\}$ are equal to $1/\tau_\upsilon$, the present MRT-LB model matches the BGK-LB model [52] for isothermal incompressible flows in porous media.

The equilibrium distribution function $f_i^{(eq)}$ for the density distribution function $f_i$ is given by

$$f_i^{(eq)} = \omega_i \left\{ \rho + \rho_0 \left[ \frac{\mathbf{e}_i \cdot \mathbf{u}}{c_s^2} + \frac{(\mathbf{e}_i \cdot \mathbf{u})^2}{2c_s^4 \varphi} - \frac{|\mathbf{u}|^2}{2c_s^2 \varphi} \right] \right\}. \tag{29}$$

where $\omega_0 = 4/9$, $\omega_{1\sim 4} = 1/9$, and $\omega_{5\sim 8} = 1/36$ are weight coefficients. The equilibrium distribution function (30) will be useful when implementing the boundary conditions.

### 3.2 MRT-LB model for the temperature field

For the temperature field governed by the energy equation (3) with a non-linear phase change source term, we propose the following D2Q5 MRT-LB equation:

$$\mathbf{g}(\mathbf{x}+\mathbf{e}\delta_t, t+\delta_t) - \mathbf{g}(\mathbf{x},t) = -\mathbf{N}^{-1}\Theta\left[\mathbf{n}(\mathbf{x},t) - \mathbf{n}^{(eq)}(\mathbf{x},t)\right] + \mathbf{N}^{-1}\delta_t \tilde{\mathbf{S}}, \tag{30}$$

where $\mathbf{N}$ is a $5\times 5$ orthogonal transformation matrix, and $\Theta$ is a non-negative $5\times 5$ diagonal relaxation matrix. The boldface symbols, $\mathbf{g}$, $\mathbf{n}$, $\mathbf{n}^{(eq)}$, and $\tilde{\mathbf{S}}$ are 5-dimensional column vectors, e.g., $\mathbf{g}(\mathbf{x},t) = (g_0(\mathbf{x},t), g_1(\mathbf{x},t), \cdots, g_4(\mathbf{x},t))^\mathrm{T}$, where $g_i(\mathbf{x},t)$ is the discrete temperature distribution function.

The five discrete velocities $\{\mathbf{e}_i \,|\, i = 0, 1, \cdots, 4\}$ of the D2Q5 model (see Fig. 2) are given by

$$\mathbf{e}_i = \begin{cases} (0,0), & i = 0 \\ \left(\cos\left[(i-1)\pi/2\right], \sin\left[(i-1)\pi/2\right]\right)c, & i = 1 \sim 4 \end{cases}, \tag{31}$$

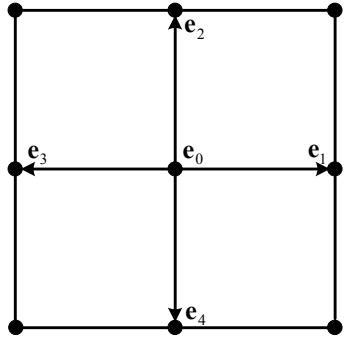

Fig. 2. Discrete velocities of the D2Q5 model.

The transformation matrix $\mathbf{N}$ linearly transforms the discrete temperature distribution functions $\mathbf{g} \in V = \mathbf{R}^5$ to the corresponding velocity moments $\mathbf{n} \in M = \mathbf{R}^5$, i.e., $\mathbf{n} = \mathbf{N}\mathbf{g} = (n_0, n_1, n_2, n_3, n_4)^\mathrm{T}$ and $\mathbf{g} = \mathbf{N}^{-1}\mathbf{n} = (g_0, g_1, g_2, g_3, g_4)^\mathrm{T}$. The transformation matrix $\mathbf{N}$ is given by [44]:

$$\mathbf{N} = \begin{pmatrix} \langle 1| \\ \langle e_x| \\ \langle e_y| \\ \langle 5\mathbf{e}^2 - 4| \\ \langle e_x^2 - e_y^2| \end{pmatrix} = \begin{pmatrix} 1 & 1 & 1 & 1 & 1 \\ 0 & 1 & 0 & -1 & 0 \\ 0 & 0 & 1 & 0 & -1 \\ -4 & 1 & 1 & 1 & 1 \\ 0 & 1 & -1 & 1 & -1 \end{pmatrix}. \tag{32}$$

The temperature $T$ is the only conserved quantity and can be calculated by

$$T = \sum_{i=0}^{4} g_i. \tag{33}$$

The equilibrium moments $\{n_i^{(eq)} \mid i = 0, 1, \cdots, 4\}$ for the velocity moments $\{n_i \mid i = 0, 1, \cdots, 4\}$ are defined as follows [50]:

$$n_0^{(eq)} = T, \quad n_1^{(eq)} = \vartheta \frac{u_x T}{\sigma}, \quad n_2^{(eq)} = \vartheta \frac{u_y T}{\sigma}, \quad n_3^{(eq)} = \varpi T, \quad n_4^{(eq)} = 0, \tag{34}$$

where $\vartheta$ and $\varpi$ ($-4 < \varpi < 1$) are parameters of the D2Q5 MRT-LB model. For $\vartheta = 1$, the MRT-LB model describes the non-linear convection-diffusion equation (3) for phase change problems. If we take $\vartheta = 0$, we can obtain an MRT-LB model for heat conduction problems with solid-liquid phase change.

The evolution process of the MRT-LB equation (30) also consists of two steps, i.e., the collision step and streaming step. The collision step is implemented in the moment space

$$\mathbf{n}^+(\mathbf{x}, t) = \mathbf{n}(\mathbf{x}, t) - \Theta\left[\mathbf{n}(\mathbf{x}, t) - \mathbf{n}^{(eq)}(\mathbf{x}, t)\right] + \delta_t \tilde{\mathbf{S}}, \tag{35}$$

the streaming step is carried out in the velocity space

$$g_i(\mathbf{x} + \mathbf{e}_i \delta_t, t + \delta_t) = g_i^+(\mathbf{x}, t), \tag{36}$$

where $\mathbf{g}^+ = \mathbf{N}^{-1} \mathbf{n}^+$.

The diagonal relaxation matrix $\Theta$ is given by:

$$\Theta = \mathrm{diag}(1, \zeta_\alpha, \zeta_\alpha, \zeta_e, \zeta_\nu). \tag{37}$$

The components of the source term $\tilde{\mathbf{S}}$ are given as follows:

$$\tilde{S}_0 = -\frac{\phi L_a}{\sigma c_{pl}} \frac{\Delta f_l}{\delta_t}, \quad \tilde{S}_1 = 0, \quad \tilde{S}_2 = 0, \quad \tilde{S}_3 = -\varpi \frac{\phi L_a}{\sigma c_{pl}} \frac{\Delta f_l}{\delta_t}, \quad \tilde{S}_4 = 0. \tag{38}$$

In simulations, the finite difference scheme $\Delta f_l / \delta_t = \left[ f_l(\mathbf{x}, t+\delta_t) - f_l(\mathbf{x}, t) \right] / \delta_t$ is employed to calculate the transient term ($\partial_t f_l$) in Eq. (3).

The equilibrium distribution function $g_i^{(eq)}$ for the density distribution function $g_i$ is given by

$$g_i^{(eq)} = \tilde{\omega}_i T \left( 1 + \frac{\mathbf{e}_i \cdot \mathbf{u}}{\sigma c_{sT}^2} \right), \tag{39}$$

where the weight coefficients $\{\tilde{\omega}_i \mid i = 0, 1, \cdots, 4\}$ are given by $\tilde{\omega}_0 = (1-\varpi)/5$, $\tilde{\omega}_i = (4+\varpi)/20$ for $i = 1 \sim 4$. The effective thermal diffusivity $\alpha_e$ is defined as $\alpha_e = \sigma c_{sT}^2 (\tau_T - 0.5) \delta_t$ with $\zeta_\alpha = 1/\tau_T$ and $c_{sT}^2 = (4+\varpi) c^2 / 10$ ($c_{sT}$ is the sound speed of the D2Q5 model).

The relationship between the liquid fraction $f_l$ and the temperature $T$ can be determined by the enthalpy method [25, 54]. In the enthalpy-based method, the interfacial position of phase change can be traced naturally through the liquid fraction $f_l$. The total enthalpy $En$, which can be split into the sensible enthalpy and the latent enthalpy, is defined by

$$En = c_p T + f_l L_a. \tag{40}$$

The relationship between the total enthalpy $En$ and temperature $T$ can be established as

$$T = \begin{cases} En/c_{ps}, & En \leq En_s \\ T_s + \dfrac{En - En_s}{En_l - En_s}(T_l - T_s), & En_s < En < En_l \\ T_l + (En - En_l)/c_{pl}, & En \geq En_l \end{cases}, \tag{41}$$

where $T_s$ and $T_l$ are the solidus temperature (temperature at the beginning of phase change) and liquidus temperature (temperature at the end of phase change), respectively; $En_s$ and $En_l$ are the total enthalpy corresponding to $T_s$ and $T_l$, respectively. The liquid fraction $f_l$ in the pore space can be calculated by

$$f_l = \begin{cases} 0, & T \leq T_s \\ (T - T_s)/\Delta \tilde{T}, & T_s < T < T_l \\ 1, & T \geq T_l \end{cases}, \tag{42}$$

where $\Delta \tilde{T} = (T_l - T_s)$. Since $f_l(\mathbf{x}, t + \delta_t)$ in Eq. (38) is not known a priori, an iterative scheme is needed in the treatment of the nonlinear latent heat source term. In this paper, the iterative enthalpy method is employed to solve the liquid fraction and temperature field. Details of the iterative enthalpy method can be found in Refs. [25, 28].

## 4. Numerical results

In this section, numerical simulations of the conduction melting in a semi-infinite space, solidification in a semi-infinite corner, and melting with natural convection in a square cavity filled with porous media are carried out to demonstrate the effectiveness and accuracy of the present MRT-LB model. In the simulations, we set $\rho_0 = 1$, $\varpi = -2$ ($c_{sT}^2 = 1/5$), $J = 1$, $\delta_t = 1$, and $\delta_x = \delta_y = 1$ ($c = 1$). The relaxation parameters $\{s_i\}$ and $\{\zeta_i\}$ can be determined by the linear stability analysis [37]. In the present work, the relaxation parameters are chosen as follows: $s_0 = s_3 = s_5 = 1$, $s_1 = s_2 = 1.1$, $s_4 = s_6 = 1.2$, $s_7 = s_8 = 1/\tau_\nu$, $\zeta_0 = 1$, $\zeta_1 = \zeta_2 = 1/\tau_T$, and $\zeta_3 = \zeta_4 = 1.5$. The non-equilibrium extrapolation scheme [55, 56] is used to treat different boundary conditions of $f_i$ and $g_i$. The MRT-LB equation of the flow field is considered for the liquid phase only, and the non-equilibrium extrapolation scheme is also employed to treat the non-slip velocity condition on the moving solid-liquid interface.

*4.1 Conduction melting in a semi-infinite space*

To validate the present enthalpy-based MRT-LB model for solid-liquid phase transition process, the one-dimensional conduction-dominated melting of a pure substance in a semi-infinite space is first simulated in this subsection. The schematic of the problem is shown in Fig. 3. Initially, the material is kept in a solid state at a temperature $T_i$ lower than the melting temperature $T_m$, then the temperature at $x=0$ is suddenly raised to $T_b$ ($T_b > T_m$) at time $t=0$ and maintained at that temperature for $t>0$. This is a two-region problem because the temperatures of the liquid and solid phases are unknown. The analytical solutions for the interface position $x_m(t)$, the liquid ($T_l$) and solid ($T_s$) temperatures are given by [57]

$$x_m(t) = 2\eta\sqrt{\alpha_l t}, \tag{43}$$

$$T_l(x,t) = T_b - \frac{(T_b - T_m)\mathrm{erf}\left[x/(2\sqrt{\alpha_l t})\right]}{\mathrm{erf}(\eta)}, \quad 0 < x \le x_m(t) \tag{44}$$

and

$$T_s(x,t) = T_i + \frac{(T_m - T_i)\mathrm{erfc}\left[x/(2\sqrt{\alpha_s t})\right]}{\mathrm{erfc}(\eta\sqrt{\alpha_l/\alpha_s})}, \quad x > x_m(t), \tag{45}$$

respectively, $\mathrm{erf}(x) = 2\int_0^x e^{-y^2} dy/\sqrt{\pi}$, and $\mathrm{erfc}(x) = 1 - \mathrm{erf}(x)$. The parameter $\eta$ can be determined from the following transcendental equation [57]:

$$\frac{e^{-\eta^2}}{\mathrm{erf}(\eta)} + \frac{k_s}{k_l}\left(\frac{\alpha_l}{\alpha_s}\right)^{1/2} \frac{T_i - T_m}{T_b - T_m} \frac{e^{-\eta^2(\alpha_l/\alpha_s)}}{\mathrm{erfc}(\eta\sqrt{\alpha_l/\alpha_s})} = \frac{\eta L_a \sqrt{\pi}}{c_{pl}(T_b - T_m)}. \tag{46}$$

In the computations, the required parameters are set as follows: $T_b = 1$, $T_m = 0$, $T_i = -1$, $T_l = 0.02$, $T_s = -0.02$, $\alpha_s = 0.02$, $\phi = 1$, $\sigma = 1$, $c_{pl} = c_{ps} = 1$, $St = c_{pl}\Delta T/L_a = 1$ ($\Delta T = T_b - T_m$). The lattice used is a $N_x \times N_y = 200 \times 6$ uniform mesh, where $N_x$ and $N_y$ are the lattice numbers along the x- and y-directions, respectively. To simulate such one-dimensional conduction melting problem, the periodic boundary condition is imposed in the y-direction, and the velocity field is set to be zero consistently ($\vartheta = 0$). In this example, $\alpha_l$ is different from $\alpha_s$ (the thermal diffusivity ratio $\alpha_l/\alpha_s$ varies from 1 to 10). The relaxation time is determined by $\tau_T = 0.5 + \alpha_e/(\sigma c_{sT}^2 \delta_t)$, where $\alpha_e = \alpha_l f_l + \alpha_s(1 - f_l)$. In Fig. 4, the temperature profiles are plotted against different values of thermal diffusivity ratio at $Fo = 0.01$ ($Fo = t\alpha_s/H^2$, here the characteristic length $H = 200$). It is seen that the MRT-LB results agree well with the analytical ones.

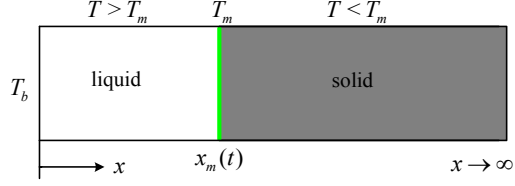

Fig. 3. Schematic of the one-dimensional conduction-dominated melting problem.

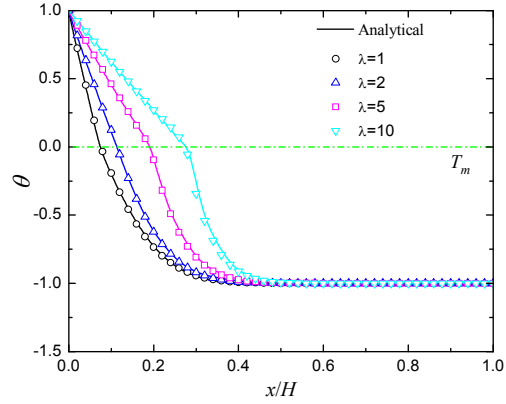

Fig. 4. Temperature profiles for different values of thermal diffusivity ratio at $Fo = 0.01$.

*4.2 Solidification in a semi-infinite corner*

In this subsection, the two-dimensional conduction-dominated solidification in a semi-infinite corner is investigated. The schematic of this problem is shown in Fig. 5. Initially, the material is kept in a liquid state at a uniform temperature $T_i$, which is higher than the melting temperature $T_m$. At time $t = 0$, the left ($x = 0$) and bottom ($y = 0$) boundary surfaces are lowered to a fixed temperature $T_c$ ($T_c < T_m$), and consequently, solidification begins along the left and bottom surfaces and proceeds into the phase change material. In order to make comparisons with the numerical and analytical results in previous studies [58, 59], we set $T_i = 0.3$, $T_m = 0$, and $T_c = -1$.

In the following simulations, the required parameters are set as follows: $T_l = 0.02$, $T_s = -0.02$, $\alpha_l/\alpha_s = 1$, $\phi = 1$, $\sigma = 1$, $c_{pl} = c_{ps} = 1$, $St = c_{pl}\Delta T/L_a = 4$ ($\Delta T = T_m - T_c$). A grid size of $N_x \times N_y = 100 \times 100$ is employed. The interface of phase change at $Fo = 0.25$ obtained by the present MRT-LB model is presented in Fig. 6 ($Fo = t\alpha_l/H^2$, here the characteristic length $H$ is chosen as $H = N_x/2$ ). The analytical solutions [58] and the numerical results obtained by Lin and Chen [59] are also presented in Fig. 6 for the purpose of comparison. It can be observed that the present numerical result agrees well with the analytical and numerical solutions. Fig. 7 shows the isothermal distributions within the semi-infinite corner at $Fo = 0.25$. As shown in Fig. 7, the gap of the isothermal line in the solid phase region is less than that in the liquid phase region. This can be attributed to the release of latent heat on the solid-liquid phase change interface. Again, the agreement with the available numerical results [59] is quite satisfactory.

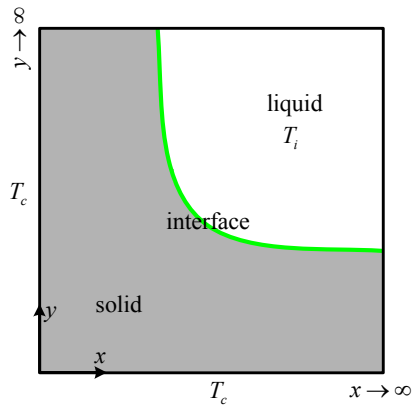

Fig. 5. Schematic of the two-dimensional conduction-dominated solidification in a semi-infinite corner.

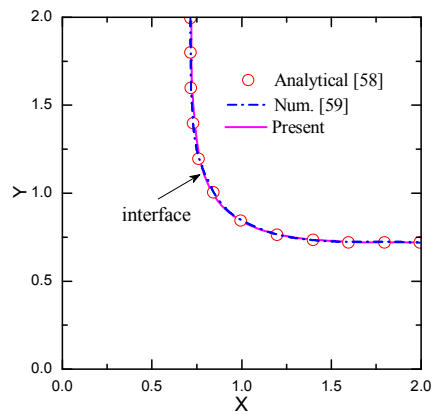

Fig. 6. Comparisons of the location of phase change interface at $Fo = 0.25$.

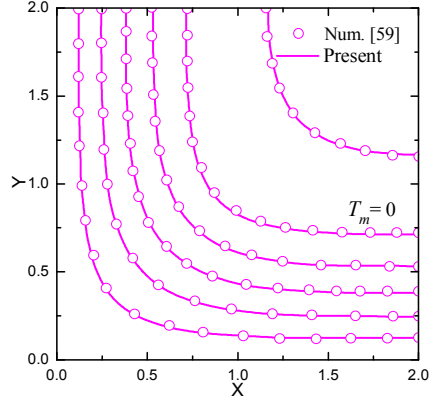

Fig. 7. Isotherms within the semi-infinite corner at $Fo = 0.25$.

### 4.3 Convection melting in a square cavity filled with porous media

In this subsection, we apply the present MRT-LB model to study the convection melting in a square cavity filled with porous media, which has been investigated both experimentally and numerically by many researchers [1, 7, 31]. The physical model and boundary conditions are shown in Fig. 8. The horizontal walls of the cavity are adiabatic, while the left and right walls are kept at constant but different temperatures, $T_h$ and $T_c$ ($T_h > T_c$), respectively. The melting temperature of the PCM is $T_m$. Initially, the PCM is in the solid state with a uniform temperature $T_i$. At time $t = 0$, the temperature of the left wall is raised to $T_h$ (higher than the melting temperature $T_m$), and consequently, melting begins along the left wall and proceeds into the PCM inside the porous cavity.

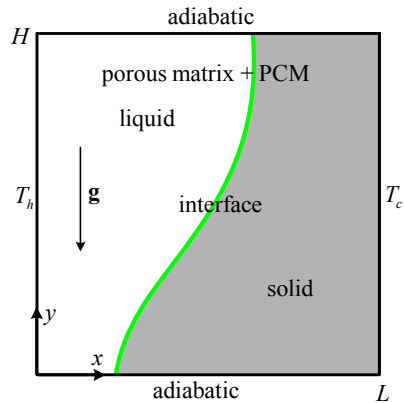

Fig. 8. Schematic of melting with natural convection in a cavity filled with porous media.

In the computations, a grid size of $N_x \times N_y = 200 \times 200$ is employed. Following Beckermann and

Viskanta [1], we set $Ra = 8.409 \times 10^5$, $Da = 1.37 \times 10^{-5}$, $\phi = 0.385$, $J = 1$, $Pr = 0.0208$, $\lambda = \alpha_e/\alpha_l = 0.2719$, $\alpha_l/\alpha_s = 1$, $T_h = 45$, $T_c = 20$, $T_m = 29.78$, $T_i = T_0 = 20$, $Fo = t\alpha_l/H^2$ (in lattice unit, $H = N_y = 200$), $St = c_{pl}\Delta T/L_a = 0.1241$ ($\Delta T = T_h - T_c$), $\sigma = 0.8604$ (in liquid region), and $\sigma = 0.8352$ (in solid region). According to Ref. [16], the dimensionless relaxation times $\tau_\upsilon$ ($\tau_\upsilon = 1/s_{7,8}$) and $\tau_T$ ($\tau_T = 1/\zeta_{1,2}$) can be determined by $\tau_\upsilon = 0.5 + \frac{MaJH\sqrt{3Pr}}{c\delta_t\sqrt{Ra}}$, and $\tau_T = 0.5 + \frac{\lambda c_s^2(\tau_\upsilon - 0.5)}{J\sigma c_{sT}^2 Pr}$, respectively, where $Ma = u_c/c_s = \sqrt{3}u_c$ is the Mach number ($u_c = \sqrt{\beta g \Delta T H}$ is the characteristic velocity, and $Ma$ is set to be 0.1 in the present study).

Fig. 9 presents the locations of the melting interface at different Fourier numbers. The experimental and numerical results given by Beckermann and Viskanta [1] are also included in Fig. 9 for comparison. It can be observed that at the early time ($Fo = 1.829$), the melting front is almost parallel to the left vertical wall, which indicates that the melting process is dominated mainly by conduction. Later in the process, the melting front gradually exhibits a shape typical for the convection-dominated melting. The melting rate decreases toward the bottom wall (the melting interface moves faster near the top wall), since the liquid cools down as it descends along the melting interface. Predictions from the present MRT-LB model are show to agree well with the results in Ref. [1]. The streamlines and dimensionless temperature fields at different Fourier numbers are illustrated in Fig. 10. The observations from Fig. 10 are in accordance with those observed in Fig. 9. To further quantify the results, the temperature profiles at different heights of the cavity for $Fo = 1.829$ and 7.314 are measured and shown in Fig. 11. It can be observed that at the early time $Fo = 1.829$ (Fig. 11(a)), the temperature profiles are close to each other because the melting is dominated mainly by conduction in the liquid region, and the temperature profiles show a variation feature of the transient evolution of the temperature field. The temperature profiles at $Fo = 7.314$ (Fig. 11(b)) indicate that the convection effect in the liquid region is somewhat stronger than that at the early time. To sum up, the numerical results of the present MRT-LB model agree well with those reported in the literature [1].

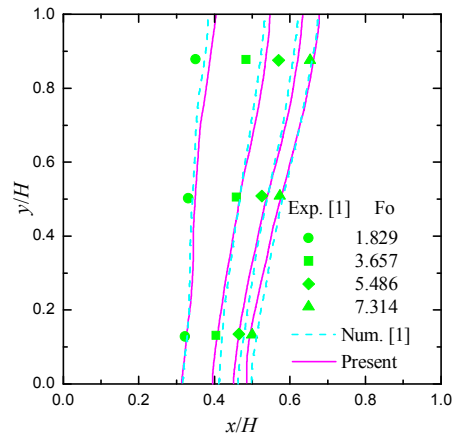

Fig. 9. Locations of the melting interface at different Fourier numbers.

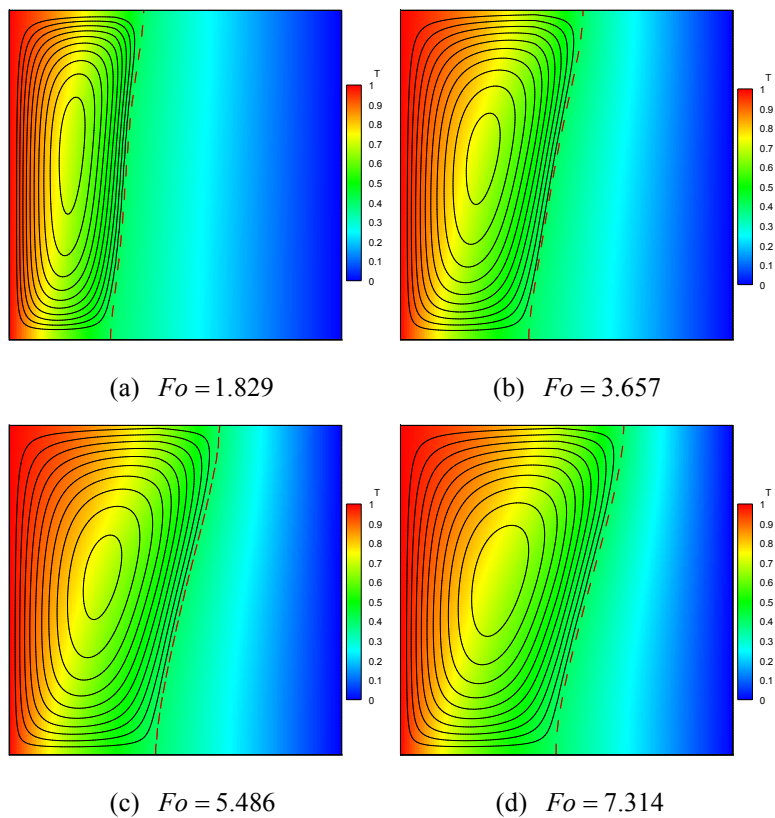

(a) $Fo = 1.829$  (b) $Fo = 3.657$

(c) $Fo = 5.486$  (d) $Fo = 7.314$

Fig. 10. Streamlines and dimensionless temperature fields at different Fourier numbers: (a) $Fo = 1.829$; (b) $Fo = 3.657$; (c) $Fo = 5.486$; (d) $Fo = 7.314$.

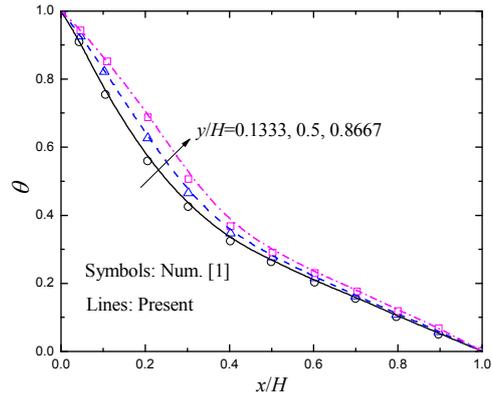

(a) $Fo = 1.829$

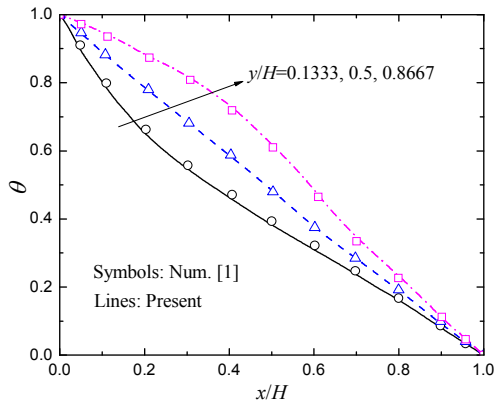

(b) $Fo = 7.314$

Fig. 11. Temperature profiles at different heights of the cavity: (a) $Fo = 1.829$; (b) $Fo = 7.314$.

## 5. Conclusions

    In this paper, a double MRT-LB model has been developed for simulating transient solid-liquid phase change problems in porous media at the REV scale based on the generalized non-Darcy model. In the model, an enthalpy-based D2Q5 MRT-LB equation is developed to solve the temperature field in addition to the D2Q9 MRT-LB equation for the flow field. The solid-liquid phase change interface is traced through the liquid fraction which is determined by the enthalpy method. Numerical simulations of the conduction melting in a semi-infinite space, solidification in a semi-infinite corner, and

convection melting in a square porous cavity are carried out to demonstrate the efficiency and accuracy of the present MRT-LB model. The numerical results indicate that the present MRT-LB model is a promising numerical technique for simulating transient solid-liquid phase change problems in porous media. Extensions of the present model for simulating solid-liquid phase change problems in porous media with variable porosity and effective properties will be reported in future studies.

**Acknowledgements**

This work was supported by the key project of National Science Foundation of China (No. 51436007) and the National Key Basic Research Program of China (973 Program) (2013CB228304).